\documentclass[letterpaper, 11pt]{article}
\usepackage{amsmath}  
\usepackage{amssymb}  
\usepackage{amsthm}  
\usepackage{enumitem}  
\usepackage{fullpage}  
\usepackage{graphicx}  
\usepackage{parskip}  
\usepackage{braket}  
\usepackage{authblk}
\usepackage{url}

\title{
    \vspace{-0.8in}
    \text{Protein folding in the modern era: a pedestrian's guide}
}

\author[1]{Sean Mullane}
\affil[1]{Department of Applied Physics, Stanford University}
\date{}

\begin{document}

\maketitle

\section{Introduction}

``Structure begets function," as the old saying goes, and for proteins this is quite literally true. The function of a protein is intimately connected to its structure, so much so that the addition of a single methyl group can make the difference between normal enzyme function and the catastrophic failure of a metabolic pathway dependent on this enzyme \cite{143LEC3}. The exquisite sensitivity of this connection fortunately lends itself to exploitation: a detailed knowledge of a protein's structure can reveal how it interacts with drugs and other proteins, accelerating drug design and disease diagnosis. A precise knowledge of how proteins fold, and the ability to predict secondary and tertiary structures from amino acid sequences, thus represents a sort of ``holy grail" for medicine and drug design. 

With that flowery (and quite frankly pretentious) introduction out of the way, I hope you're excited to join me for this discussion of protein folding! While lots of methods have developed since the eighties to tackle the \textit{protein folding problem} and \textit{protein structure prediction}, the last five years or so have seen the rise of two particularly powerful and exciting methods: machine learning and quantum computation. I'll walk through the basic concepts motivating each approach, and try to provide a summary of current state-of-the-art research in these fields. I've structured this paper like a textbook chapter, aiming for a pedagogical approach over a lightning-fast review filled with jargon and equations. Please enjoy! 

\section{Folding: basic concepts}

To begin a brief review of fundamental concepts in protein folding, I want to state what is perhaps the field's most surprising result: 

\textit{The final structure of a protein (secondary and tertiary structure) is completely encoded in its primary amino acid sequence.}

This is \textit{astounding}. Given that chaotic systems are so abundant in nature, one might expect that small deviations in the translated amino acid chain (for example, deviations in the relative positions of residue side chains due to thermal fluctuations) would lead to a vast array of mostly non-functional secondary and tertiary structures. However, this is emphatically not the case; clever experiments performed by Christian Anfinsen and others (which we'll describe soon) showed that only the primary amino acid sequence is necessary to reach the functional protein structure \cite{NOBEL}. This is ultimately a consequence of Anfinsen's \textit{thermodynamic hypothesis}, which states that the final, functional structure of a protein (also called the ``native" structure) is the structure that minimizes the Gibbs free energy $\Delta G$.

While Anfinsen explored the thermodynamics of folding, we find equally interesting puzzles in an exploration of folding \textit{kinetics}: once the primary amino acid chain is specified, proteins tend to fold extremely quickly. This observation led Cyrus Levinthal to propose a question that has since become known as \textit{Levinthal's paradox} \cite{PABLO}:

\textit{How do proteins fold so quickly when their associated conformational spaces can be astronomically large?}

In this context, ``conformational space" means the set of all possible conformations for a given amino acid sequence. As an example, suppose we have a 100-amino acid polypeptide where each peptide bond is described by the two dihedral angles $\phi$ and $\psi$ (we'll assume the \textit{trans} conformation for the amide bond, for simplicity). If we restrict the angles to the ranges $-60^{\circ} \leq \phi \leq -120^{\circ}$, $60^{\circ} \leq \psi \leq 120^{\circ}$ (typical values for a $\beta$ sheet \cite{141LEC4}) and suppose for simplicity that the angles only take on integer values, there are 60 possible values for each angle \textit{per peptide bond}. This gives $2 \cdot 60^{100} \approx 10^{178}$ possible conformations of the polypeptide! As noted in \cite{PABLO}, randomly searching through all of these states would take far longer than the life of the universe. Thermodynamics might tell us where the protein's native structure exists (at the minimum $\Delta G$), but it doesn't tell us how to get there, a problem which has led to much investigation into protein folding kinetics.

In the rest of this section, we'll take a closer look at Anfinsen's experiments, and how they lend support to the thermodynamic hypothesis and the idea that the amino acid sequence encodes all necessary information for the protein's native structure. Then we'll complete a brief review of folding kinetics, looking in particular at the shape of the protein folding free-energy landscape and how these landscapes offer a solution to Levinthal's paradox. It turns out that understanding free-energy landscapes will be critical to our analysis of machine learning and quantum computing methods for predicting how proteins fold!  

\subsection{Thermodynamics: Anfinsen's experiments}

\begin{figure}
    \centering
    \includegraphics[scale=0.5]{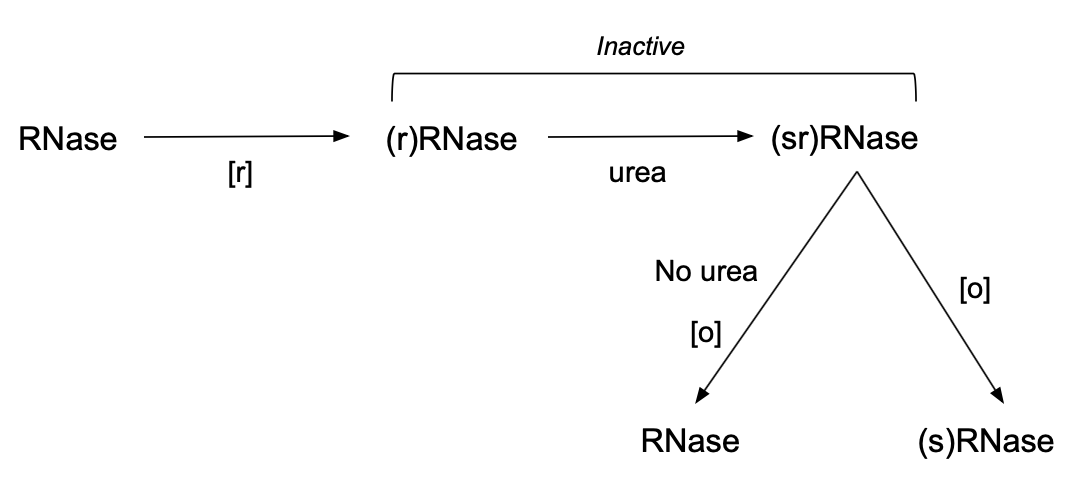}
    \caption{The overall experimental process Anfinsen and colleagues used in their experiments, in particular for \cite{1961}. Diagram is very loosely based on figure 3.1 in \cite{PABLO}.}
    \label{anfinsen_exp}
\end{figure}

\begin{figure}
    \centering
    \includegraphics[scale=0.5]{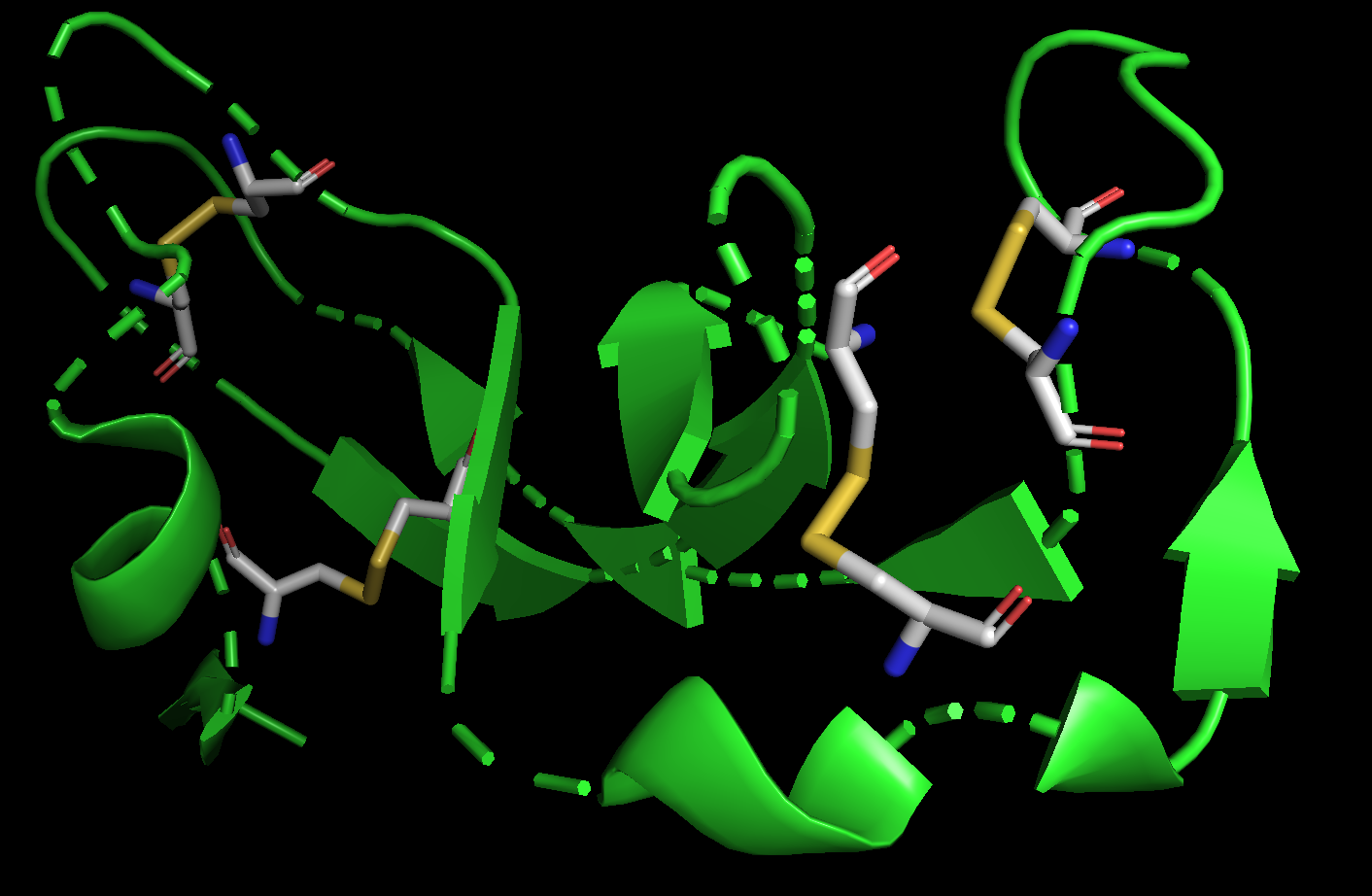}
    \caption{The local environment of the four bovine RNase disulfide bridges, which are highlighted in gold. The green structure does not represent the entire protein; I've only included residues within 5 \AA \ of the disulfide bridges for clarity. Only this arrangement of disulfide bonds (out of 105 total possibilities) leads to a functional enzyme.}
    \label{rnase}
\end{figure}

Over the course of 1960 -- 1965, Anfinsen and various colleagues completed a remarkable series of investigations and published at least six papers that ultimately culminated with the thermodynamic hypothesis and a Nobel prize for Anfinsen. To showcase his experiments and logic we'll focus on \cite{1961}, which summarizes the elegant methods he used to determine that the primary sequence encodes the necessary information for protein folding. 

Anfinsen's experiments focused on bovine ribonuclease, a 13.7 kDa/124 residue RNase that cleaves single-stranded RNA strands (PDB 1FS3; see figure \ref{rnase}). This enzyme was useful for protein folding experiments because the native form contains four disulfide bonds between cysteine residues that can be easily reduced (to form eight --SH groups) and re-oxidized \cite{1960}. With eight total cysteine residues, there are $7 \cdot 5 \cdot 3 = 105$ possible permutations of S--S bonds, but only \textit{one} of these permutations results in a functional enzyme (we'll describe how Anfinsen determined this below). Anfinsen thus had significant control over the structure of the enzyme by simply manipulating the disulfide bridges. 

His experimental setup is shown in figure \ref{anfinsen_exp}. Beginning with functional ribonuclease (``RNase"), he added a reducing agent (generally mercaptoethanol) to affect the S--S $\rightarrow$ SH $+$ SH transition and a ``scrambling" or ``denaturing" agent (urea) to further disrupt the secondary and tertiary structure of the protein. The final products from these reactions are indicated by (r)RNase = reduced-RNase and (sr)RNase = scrambled-reduced-RNase in the diagram. No enzymatic activity was found by Anfinsen upon adding substrate to the (r)RNase or (sr)RNase solutions; since both mercaptoethanol and urea significantly disrupt the native secondary and tertiary structures, this result wasn't surprising. 

As a next step, Anfinsen removed the (sr)RNase from the urea solution, adjusted the pH, and exposed the resulting solution to oxygen in order for the SH $+$ SH $\rightarrow$ S--S oxidation to occur. At regular intervals during this reoxidation process, Anfinsen performed the following experiments: 
    \begin{enumerate}
        \item \textbf{Carbon labeling.} The solution was mixed with labeled iodoacetic acid; any cysteine residues in the reduced form could displace iodine to form radiolabeled-alkylated cysteine. After about 20 hours of reoxidation, the amount of radiolabeled cysteine detected in the solution almost exactly matched the amount detected in the native enzyme solution\footnote{Anfinsen didn't report absolute values of the radiolabeled cysteine in his paper; instead, he reported results as ``percent of maximal change", equivalent to (100\% -- percent difference between the values in experimental solution and control solution). After 20 hours of oxidation, this value was about 95\%.}. 
        \item \textbf{Optical activity.} Anfinsen compared the optical activity (amount of rotation of plane-polarized light) of the solution and native protein, and found excellent agreement between the rotations after 20 hours of oxidation (\% of maximal change $\gtrsim 95\%$).
        \item \textbf{Absorption spectrum.} The (sr)RNase sample absorption spectrum was peaked at about 275 $\mu$m, while the native RNase sample spectrum peaked at about $277.5 \ \mu$m. During reoxidation, the experimental sample's peak moved steadily upwards from 275 $\mu$m, and eventually settled at exactly $277.5$ $\mu$m after about 20 hours of oxidation. 
    \end{enumerate}
These results strongly suggest that the reoxidation process was able to almost completely re-form the native enzyme structure, even after breaking all the sulfide bonds \textit{and} scrambling the secondary and tertiary structures! I really want to emphasize how astounding this is--aside from exposure to oxygen, nothing in the solution was present to assist the proteins with folding. Not only that, but out of the 105 possible combinations of sulfide bonds, \textit{almost every protein in solution was able to form the single sulfide bond combination that results in a functional enzyme}.

``That's great, Sean", you may ask, ``but how do we know that only 1 of the 105 combinations results in a functional enzyme? These results would be a lot less impressive if most of the combinations were acceptable for enzyme function." Great question, dear reader! This brings us to one of Anfinsen's corollaries to this experiment: after forming a second solution of (sr)RNase, he began the reoxidation process (by exposure to oxygen) \textit{without} removing urea from solution. This produced a solution, denoted by (s)RNase = ``scrambled-RNase" in figure \ref{anfinsen_exp}, where enzymatic activity was only $\approx 1\%$ that of the native protein solution \cite{NOBEL}. Since urea is a denaturing agent, it's unsurprising that the vast majority of the proteins were not able to fold into a functional form. What's more surprising is that a small fraction of enzymes \textit{were} able to fold to functional forms, and that this ``small fraction" works out to about $0.01 \approx 1/105$. Hopefully you see where I'm going with this; let's follow the logic through carefully. If no enzymatic activity is detected in the (r)RNase solution (reduced, but not denatured), we know that disulfide bonds are necessary for enzyme function. Performing an oxidation reaction to obtain the (s)RNase solution constructs these disulfide bonds, but only $1\%$ of the enzymes in this solution are actually functional. If we assume that most or all of the enzymes were completely oxidized (easily checked via the carbon labeling method), this means that only about $1\%$ of the possible disulfide bridge pairings lead to a functional enzyme--about 1 in 105!  

Just based on the results of Anfinsen's simple experiments, we're able to conclude with reasonable certainty that only a single disulfide bridge combination leads to a viable enzyme, and that a denatured protein mixture can ``find" this disulfide pairing to produce completely functional enzymes. Anfinsen took this as strong evidence for the thermodynamic hypothesis we presented earlier: if (sr)RNase can find a native form in a test tube, with nothing else present (aside from exposure to oxygen), the free  energy $\Delta G = \Delta H - T\Delta S$ is the only possible quantity that could drive folding. Although the entropy $\Delta S$ tends to \textit{decrease} during folding, as the unfolded chain has far more available conformations than the functional enzyme, energy-lowering interactions encoded in $\Delta H$ (like hydrogen bonding and the hydrophobic effect) compensate to make $\Delta G < 0$. As we'll see soon, this idea of minimizing the free energy to drive folding is incredibly powerful when we develop algorithms to predict protein structures. 

Before moving on, I want to make a brief note about a puzzle that the discerning reader might recognize: it took many hours (typically $> 20$) for the (sr)RNase to fold into functional RNase during Anfinsen's experiments. An organism generally can't wait 20 hours for critical proteins to be folded, so why is the process so slow in the test tube? It turns out that an enzyme exists in the bovine endoplasmic reticulum that catalyzes the formation of the disulfide bonds during folding \cite{1963}. Adding purified samples of this enzyme to the (sr)RNase solution accelerated the folding process to less than two minutes, a typical value observed \textit{in vitro} for this protein \cite{NOBEL}. Note in particular that this does \textit{not} negate the hypothesis that only the amino acid sequence is necessary for folding. Without the enzyme, the protein still folds; the ER enzyme only affects the \textit{kinetics} of the reaction, and does nothing to change the thermodynamics (just as we expect from our basic notions of enzymes in the body).  

\subsection{Kinetics}


I want to briefly return to Levinthal's paradox and the notion of kinetics before we begin discussing machine learning methods. Since we now know that the protein's native structure is found at the minimum of its free energy function, it's natural to ask exactly what this function looks like and how its form affects the speed of folding. This, in a nutshell, describes folding kinetics: since the protein starts in a higher-energy state, what path along the ``free energy landscape" does it take to reach the minimum $\Delta G$? If that path leads directly to the global minimum, the protein will fold quickly; if the path leads through lots of \textit{local minima}, it might take the protein a long time to find the minimum energy state, or might keep the protein from finding that state altogether\footnote{For those unfamiliar with optimization, a \textit{global} minimum is the point at which $\Delta G$ is smallest (most negative) for \textit{all} possible inputs to the function. The \textit{local} minimum is a point where $\Delta G$ is the smallest, but only for some finite region around that point.}. How might this idea of a ``path" through the free energy landscape let us resolve Levinthal's paradox? If a protein's free energy landscape is \textit{funneled}, so that all initial higher-energy states will ``roll" towards the global minimum, it's reasonable to assume that folding could be accomplished quickly. This idea is nicely illustrated in figure 5 of \cite{REVIEW}; note how a funnel structure allows us to converge to the optimal structure much faster than if we had to randomly search for a very narrow region in which $\Delta G$ is small. Of course, the free energy function for an actual protein would be impossible to visualize, as it would depend on many parameters and exist in more than three dimensions, but these toy models should help us build a conceptual picture of folding kinetics. Indeed, the current prevailing theory is that free energy functions are roughly funnel-shaped, a property that is sometimes exploited to develop folding algorithms \cite{REVIEW}. This helps us answer Levinthal's paradox: nature selected proteins with funnel-shaped free energy functions, so fast folding was practically guaranteed to occur. Any slow-folding proteins with difficult-to-find global minimums provided no evolutionary advantage, and were selected out of the population. 

It's worth noting that different proteins can still fold over vastly different timescales, so the above discussion is a bit of an over-simplification; for those looking to read more into folding kinetics, \cite{REVIEW} is an excellent place to start. However, this is what we'll need to know in order to explore machine learning and quantum computing based methods in the next few sections. 

\section{Machine learning}

Okay, a normal introduction to this section would spend some time trying to convince you why machine learning is useful, but by this point I hope you're already convinced of that, given that you probably use an iPhone or Pixel every day. While I can't say I'm a huge fan of machine learning methods (for reasons I'll discuss soon), no one can doubt their usefulness, and the protein folding problem is an area where they might be especially useful. Much effort has been invested into developing machine learning models for protein folding; the review \cite{ML} is a great place to start if you're interested. However, I thought we'd begin at the very top and discuss a recent model that has outperformed most others: the AlphaFold algorithm from a team at DeepMind. 

\textit{(Sidenote: If you're not familiar with some basic machine learning concepts, the next section might be a little difficult. I'm not making it too technical, but it will be helpful to know the basic ideas behind neural networks and be able to recognize the following terms: convolutional network, histogram, gradient descent, backpropagation, tensor, and loss function. If you want a good introduction to neural networks, google ``Stanford CS231N" and check out their lecture notes!)}

\subsection{AlphaFold: Background}

AlphaFold is a deep neural network that competed in the 2018 CASP protein structure prediction competition, and was designed to excel at \textit{free-model protein structure prediction}. The competition's other category, template-based modeling, involves predicting the structure of a protein with an amino acid sequence that is very similar (or \textit{homologous}) to at least one other known protein structure. These algorithms begin with the known structure, and make changes according to differences in the amino acid chains to arrive at the unknown protein's structure. In free-model prediction, no such homologous structure is known; the algorithms must predict the native protein structure from only the amino acid sequence and \textit{multiple sequence alignment data} \cite{ALPHA}. We'll discuss the details of the input data later in this section, but it's important to recognize now that free-model prediction is much harder than template based modeling.

Let's start with the basic idea behind the algorithm: 

\textit{If we can construct a protein-specific potential function, we can perform gradient descent to minimize this function and obtain the protein's native structure.}

In this case, a ``potential function" (denoted by $V$) is equivalent to the $\Delta G$ functions we discussed in the last section; it very loosely describes interactions among the protein's residues, so high-energy interactions increase $V$ and energy-lowering interactions decrease $V.$ As we'll see, this potential function is not constructed by considering hydrogen bonding and entropy explicitly, but by constructing probability distributions output from a neural network. A minimization of $V$ thus corresponds to a protein with the most possible energy-lowering interactions, equivalent to a minimization of $\Delta G$ to determine the native structure in the previous section.   

Minimizing $V$ might correspond to the highest number of energy-lowering interactions, but how does this help us actually determine the protein's structure? We can equivalently think of $V$ as being a function of the protein \textit{conformation}, since high-energy conformations correspond to high-energy interactions. Any protein conformation can by parameterized by a set of dihedral angles $\{ \phi_i, \psi_i \}$ for each residue, so it seems natural to require that our constructed potential $V$ is a function of these dihedral angles: 
    $$ V = V(\{ \phi_i, \psi_i \}) $$
If we can determine the set of dihedral angles $\{ \phi_i, \psi_i \}_{\text{min}}$ that minimizes $V$, we'd know the dihedral angles that give the most energy-lowering interactions and thus have an almost complete description of the native structure. Determining $\{ \phi_i, \psi_i \}_{\text{min}}$ can be accomplished by running a gradient descent algorithm on $V$. For the mathematically inclined, the AlphaFold algorithm is just solving the minimization problem: 
    $$ \min_{\phi, \psi \in [0, 2\pi]} V(\{ \phi_i, \psi_i \}) $$
Of course, all of this is contingent on being able to construct the function $V$ accurately! The rest of the AlphaFold algorithm is designed to construct $V$, based \textit{only} on the primary amino acid sequence and multiple sequence alignment (MSA) data, which we'll discuss below. 
\subsection{AlphaFold: Constructing the potential}

The function $V$ is generated by a deep convolutional neural network, with an input based on the amino acid sequence and MSA data. I'm not going to spend too much time on the group's processing of the input data, as it's fairly complex and not necessary to understand the paper's important ideas, but I do want to briefly discuss the MSA data. Although free-modeling algorithms can't \textit{begin} with a homologous protein structure and make edits according to differences in amino acid sequences, they can search databases for sequences similar to the unknown protein's sequence. This is called \textit{multiple sequence alignment}, and in its most basic form entails lining up similar amino acid sequences and looking for insertions, deletions, and shifts of residues \cite{MSA}. Intuitively, we expect similar sequences to have relatively similar native structures, and DeepMind researchers wanted AlphaFold to factor these similar sequences into its construction of the potential. Through a lot of preprocessing, they were able to construct an input matrix dependent on sequence alignment data that could be processed by a deep convolutional network. It's important to note that the network does not explicitly analyze sequence correlations like a normal MSA algorithm; instead, the researchers constructed a \textit{representation} of this data, and trained the network to use this representation to construct $V$.

Let's focus next on the \textit{output} of the deep convolutional network, as this is incredibly important in understanding how the function $V$ is ultimately created. The convolutional network outputs a discrete probability distribution over the distances between residues, denoted by $\mathcal{P}(d_{ij} | S, \text{MSA})$. This can be read as, ``The probability that the distance between residue $i$ and residue $j$ is $d_{ij}$, given the amino acid sequence $S$ and the MSA data." The network also outputs a discrete distribution over the dihedral angles for each residue: $\mathcal{P}(\phi_i, \psi_i | S, \text{MSA})$, or ``The probability that the dihedral angles for residue $i$ are $\phi_i$ and $\psi_j$, given $S$ and the MSA data." How is this data represented? For the distance distribution, the engineers consider residues that were within 22 \AA \ of one another, and split the range 2 -- 22 \AA \ into 64 equal bins. For a given bin, they constructed an $L \times L$ matrix, where $L$ is the total number of residues in the protein. Entry $(i, j)$ in this matrix represents the probability $\mathcal{P}$ that residue $i$ is within a distance $x$ of residue $j$, where $x$ is the upper bound of the range defined by the given bin. For example, given the $2 - 2.33$ \AA \ bin, entry $(1, 2)$ represents the probability that residue 1 and residue 2 are within $d_{12} = 2.33$ \AA \ of each other. Since there are 64 total bins, the output of the network is an $L \times L \times 64$ dimensional tensor representing the discrete probability distribution $\mathcal{P}(d_{ij} | S, \text{MSA})$, since the output tensor is conditioned on the input amino acid sequence and MSA data. The authors call this tensor a ``distogram," short for ``distance-histogram," as it represents a histogram over possible inter-residue distances within the protein. The dihedral distribution was constructed similarly, with discretized angle bins and an output tensor representing $\mathcal{P}(\phi_i, \psi_i | S, \text{MSA})$. Note that this distribution is simpler because it doesn't rely on values between residues, so the output tensor is only $L \times 1,296$ dimensional ($1,296$ bins were used to discretize the angles).

What do we do with these probability distributions, now that we have them? To answer this, I want to first consider how the researchers formulated the potential $V$. The overall function consists of three terms: a \textit{distance} term that takes as input the distance distribution, and \textit{angular} term that takes as input the angular distribution, and a \textit{van der Waals} term that penalizes steric interactions between atoms\footnote{The van der Waals term was previously developed by the Rosetta software package, so its explicit form wasn't discussed in the paper.}. So we have: 
    $$ V(\mathcal{C}) = V_{\text{dist}}(\mathcal{C}) + V_{\text{angle}}(\mathcal{C}) + V_{\text{vdw}}(\mathcal{C}) $$
Here we write $\mathcal{C}$ to represent that each $V$ above is a function of the conformation of the protein. But we already know that the conformation can be parameterized by the dihedral angles, so: 
    $$ V(\{ \phi_i, \psi_i \}) = V_{\text{dist}}(\{ \phi_i, \psi_i \}) + V_{\text{angle}}(\{ \phi_i, \psi_i \}) + V_{\text{vdw}}(\{ \phi_i, \psi_i \}) $$
Let's take a look at the form of each term. The distance potential was written as: 
    $$ V_{\text{dist}}(\{ \phi_i, \psi_i \}) = - \sum_{i, j | i \neq j} \log{\mathcal{P}(d_{ij} | S, \text{MSA})} $$
The angular potential was written similarly: 
    $$ V_{\text{angle}}(\{ \phi_i, \psi_i \}) = - \sum_{i} \log{\mathcal{P}(\phi_i, \psi_i | S, \text{MSA})} $$
Why are these useful? Suppose we \textit{maximize} the distance distribution, so we find the distance $d_{ij}$ that gives the highest probability $\mathcal{P}$ between residues $i$ and $j$. Maximizing a non-zero function is equivalent to \textit{minimizing} its negative log, which is exactly what each potential above does. So if we minimize $V_{\text{dist}}$ and $V_{\text{angle}}$, we're essentially just finding the distances $d_{ij}$ and angles $\phi_i, \psi_i$ that have the highest probability of existing in the actual protein's native structure. In practice, we'd write the distances $d_{ij}$ in terms of the dihedral angles, as inter-residue distances are directly dependent on the set of dihedral angles in the protein. This means that when minimizing our overall potential $V$, we're just looking for an assignment of angles $\{ \phi_i, \psi_i \}$ for all $i$ that leads to the smallest possible value of $V$. Once these angles are determined, we effectively have a complete representation of the protein's backbone, from which we can reconstruct the secondary and tertiary structures! 

\textit{(As a slight technical aside, you might notice that the convolutional network constructed a discrete distribution over the distances and angles, but in order to use stochastic gradient descent we must have a continuous, differentiable function. To fix this, the engineers fit a spline to the negative log of the discrete probability distribution, which constructed a differentiable function.)}

\subsection{AlphaFold: Results and discussion}

The paper brags in some detail about the algorithm's performance at CASP13, so I'll just summarize their results here. They did well! Very well, in fact. The CASP competition has a scoring system called the \textit{template modeling} (or TM) score, which uses a complex metric to quantify the degree of overlap between an algorithm's proposed structure and the protein's actual native structure. Out of the six brand new proteins used in the CASP13 competition, the AlphaFold algorithm outperformed all other groups for five proteins (see figure 1 in \cite{ALPHA}). I don't want to spend too much time discussing other performance metrics of the algorithm, as you can read their paper for a more detailed analysis of their results. Instead, I want to briefly comment on the philosophy of their approach to protein folding. Neural network models are often called ``black box" models, because we don't really know ``how" the neural network actually learns to produce the output distribution over distances and angles. Of course, the overall idea is fairly simple: during training of the neural network, the output distribution is compared to some ground-truth distribution, and any error is backpropagated to the network's weights. But gaining insight into exactly \textit{how} this network functions when deployed--which weights are important for a given input feature, for example--is extremely difficult, as trained neural networks are highly complex nonlinear functions. Instead of relying on physics to construct the potential function, we allow the convolutional network to draw inferences from the input data and build our potential from there. 

This isn't necessarily a bad thing; techniques that try to construct potentials from the physics of residue interactions are often highly complex and difficult to simulate. A perfectly accurate ``hand-constructed" potential would have to take into account the interactions of any given atom with all other atoms in the system; even for a moderately sized protein with 100 atoms, this would result in $99! \approx 10^{155}$ interactions! Since there are $\approx 10^{50}$ atoms in the entire Earth, it's unlikely that we'd ever develop systems large enough to fully simulate these interactions. Physicists get around this by making clever approximations when constructing potentials, like noting that van der Waals interactions and hydrogen bonds are very short-ranged. Even with these approximations, however, physics-based potentials can still hit the limit of classical computing power very quickly. In some sense, using a neural network abstracts away this complexity and allows us to construct potentials that are good enough for the proteins in the CASP database ($\approx 500$ residues or less). Still, as a physicist I'd be interested to see if AlphaFold could be complemented with a physics-based approach to improve performance (this already happened in a small sense, with the extra van der Waals term added to the potential function). This might be an especially relevant question when it comes to \textit{scaling}; because the paper \cite{ALPHA} focused so much on the CASP competition, the authors mention nothing about how their algorithm would perform on other proteins (for example, those with $\approx 10^3$ residues). I'd be interested to see if the model's performance would scale with the protein's size and complexity, and whether a complementing physics-based model might improve performance in this regime.  

\subsection{Alternative approaches}

I want to briefly contrast the method used by the AlphaFold team and discuss the approach used by \cite{HARVARD} to build a protein-prediction algorithm. As we noted above, AlphaFold used a neural network to construct a probability distribution, from which they derived an energy function to find the native structure. It turns out that you can also take the \textit{opposite} approach and attempt to construct a probability distribution over all states from a learned energy function. This is how \cite{HARVARD} approaches the protein-folding problem; they noted from statistical mechanics that the probability of a thermodynamic system (in this case, our protein) being in a conformation $\mathcal{C}$ is given by: 
    $$ \mathcal{P}(\mathcal{C}) = \frac{e^{-V(\mathcal{C})}}{Z} $$
This is called the \textit{Boltzmann distribution}; note that $V(\mathcal{C})$ is the energy function we considered in the previous section (it might not be identical in form to the AlphaFold team's function, but it encodes similar information). For the mathematically inclined, I've included a brief addendum below on what the variable $Z$ is doing in this equation, but it's not super important to understand for our purposes. We just need to understand why it's useful to have a probability distribution over all the conformations: if we can \textit{sample} from this probability distribution, we can find the conformation with the highest probability, and assign that as our native protein structure. Note that having a high probability of some conformation $\mathcal{C}$ corresponds to a \textit{low} energy for this conformation, because there is a minus sign in front of $V$ in the exponential. This approach is essentially the same as the AlphaFold process, but chooses not to directly minimize the energy function. 

In \cite{HARVARD}, the authors use a neural network to construct the energy function $V(\mathcal{C})$, again using a representation of the primary amino acid sequence and data from similar amino acid sequences as input. They then use a type of stochastic differential equation called \textit{Langevin dynamics} to construct and sample from the Boltzmann distribution we mentioned above, and use these samples to ultimately construct a protein. I don't want to spend any time discussing their exact process, as it's highly technical and would require a significant extension to this already lengthy paper! Instead, I'd like to comment on how this approach is somewhat more physics-based than the AlphaFold team's approach. The Boltzmann distribution they use is a fundamental object in statistical mechanics, where it's used to compute the properties of large collections of molecules (for example, an ideal gas). Although \cite{HARVARD} relies heavily on machine learning to actually construct the functions they need, the entire approach is grounded in an assumption that the protein system should obey the laws of statistical mechanics. This assumption was absent from the process used by AlphaFold, which only assumed that the native protein structure could be found at the minimum of some energy function. Again, I'm not necessarily arguing that this physics-based approach is better, especially since AlphaFold currently outperforms almost all other competitors. I just want to highlight how these two representative papers take different philosophical approaches to the computation of protein structure: we can let a neural network learn everything on its own, or we can supplement the neural network with our knowledge of how dynamical systems behave. Both approaches, I suspect, will be important in the coming decades as protein folding becomes even more useful for drug design and disease diagnosis! 

In the next section, we'll see this idea of physics-based methods taken to its extreme with the use of \textit{quantum computers}, which are themselves highly complex quantum-mechanical systems, to simulate protein folding! 

\textit{(Aside: I promised an explanation of the factor of $1/Z$ appearing in the Boltzmann distribution above. For a complete explanation please consult your friendly neighborhood statistical mechanics textbook, but in brief: the quantity $Z$, which is called the partition function, ensures that the probability distribution is normalized, so the probabilities over all conformations add up to one. The partition function is defined as: 
    $$ Z = \frac{1}{N! h^{3N}} \int (d^{3N}p) (d^{3N}x) \ e^{-V} $$
where $N$ is the number of particles, $x$ and $p$ are the position and momenta of each particle, and $V$ is the energy function \cite{STAT}. Defined in this way, we ensure that integrating the quantity $\mathcal{P}(\mathcal{C})$ over all possible conformations gives
    $$ \int d\mathcal{C} \ \mathcal{P}(\mathcal{C}) = 1 $$
as we require for a probability density function. Note that $Z$ is essentially impossible to analytically compute for all but the simplest of systems, which is why \cite{HARVARD} uses an approximation (Langevin dynamics) to actually sample from $\mathcal{P}$. They avoid the near-impossible task of computing $Z$ for a large protein!)
}

\section{Quantum computing}

Why is quantum computing useful for protein folding? To explain this, we'll first have to explain the basic concepts behind quantum computing! I'll assume here only a basic knowledge of linear algebra, so you should be comfortable working with vectors and matrices. There are many excellent resources out there if you want to learn more about the theory behind quantum computing; the canonical textbook is \cite{MIKEIKE}, although it assumes that you've taken some undergraduate-level linear algebra and computer science classes. 

The basic unit of a quantum computer is the \textit{qubit}, which is a portmanteau of ``quantum bit." Just like a classical bit, a qubit can be in one of two states: the ``0" state, which we denote mathematically by $|0\rangle$, and the ``1" state, which we denote by $|1\rangle$. The funny brackets around the 0 and 1 are from a specific type of physics notation known as \textit{bra-ket notation}, but for our purposes the details aren't super important. You can just think of $|0\rangle$ and $|1\rangle$ as representing the two possible states of a classical bit. How are qubits different? It turns out that a qubit exists in a \textit{superposition} of these two states, instead of ``only" being in the 0 state or ``only" being in the 1 state. If we represent our qubit by $|\psi\rangle$, we represent this superposition mathematically as: 
    $$ |\psi\rangle = \alpha |0\rangle + \beta |1\rangle $$
This equation looks a little weird when you first see it, so let's break it down carefully. The $|\psi\rangle$ term on the left hand side is just how we represent a qubit; the use of the Greek letter $\psi$ and the brackets have origins in physics, but again the details aren't important right now. Whenever you see $|\psi\rangle$, you can mentally replace it with ``the state of the qubit." We've already noted that $|0\rangle$ and $|1\rangle$ represent the 0 and 1 states that correspond to classical bit states. The variables $\alpha$ and $\beta$ above are \textit{complex numbers} that define the superposition over the 0 and 1 states. Why are these numbers necessary? If we have a qubit in a quantum computer, we can \textit{measure} the qubit and try to observe its state. If we do this, we'll always find that the qubit is \textit{either} in the $|0\rangle$ state or the $|1\rangle$ state. The numbers $\alpha$ and $\beta$ define \textit{how often} we'll observe the $|0\rangle$ and $|1\rangle$ states, respectively, if we have many copies of the same qubit and perform a measurement on each one. To make this more precise, the \textit{probability} of observing the $|0\rangle$ state is $|\alpha|^2 = \alpha^*\alpha$, while the probability of observing $|1\rangle$ is $|\beta|^2 = \beta^*\beta$ (here, the $^*$ operation denotes complex conjugation). Since upon measurement we must observe either $|0\rangle$ or $|1\rangle$, the probabilities of obtaining these states must sum to one: $|\alpha|^2 + |\beta|^2 = 1$. A simple example is the following state: 
    $$ |\psi\rangle = \frac{1}{\sqrt{2}} |0\rangle + \frac{1}{\sqrt{2}} |1\rangle $$
If we measure this qubit, the probability of obtaining $|0\rangle$ is $|(1/\sqrt{2})|^2 = 1/2$, and we clearly have the same probability of obtaining $|1\rangle$. So we have a 50\% chance of observing either state, similar to a fair coin flip. 

I want to briefly comment on how to interpret these results, as there's a strong tendency to try and relate these concepts to things we're familiar with (like the coin flip) when first learning them. Popular science articles discussing qubits will usually say something along the lines of, ``the qubit simultaneously exists in both the $|0\rangle$ and $|1\rangle$ states, and then picks one of these states when we try to observe it." While this interpretation seems reasonable given what we've discussed above, it's actually \textit{wrong}. A more precise way to describe the qubit is something along the lines of: \textit{the qubit \textbf{does not} have a well defined state before you observe it; only after observation does the qubit ``collapse" into the $|0\rangle$ or $|1\rangle$ state.} We could spend hundreds of pages trying to dissect exactly what this interpretation means, but the bottom line is that qubits are quantum mechanical objects that are \textit{utterly different} than anything human beings interact with in our normal lives. Quantum mechanics requires so much mathematics because there is no way to precisely describe concepts like a qubit with modern language. For the purposes of this paper, just remember that terms like ``the qubit is in a superposition of 0 and 1" are imperfect ways to represent highly complex mathematical objects with language; take these analogies with a grain of salt! 

Okay, back to protein folding: why do we want to use quantum computers to predict protein structures? One of the most compelling reasons connects back to the very beginning of this paper, when we noted how a typical polypeptide can have something on the order of $10^{178}$ possible conformations. This means algorithms predicting conformations have to make some significant assumptions to restrict the size of the conformation space they're searching, and even after these assumptions the algorithms can still take days to train and ultimately produce results. It turns out that using quantum computers gives us access to a very large space in which to compute, far larger than any classical computer could ever hope to achieve. Here's the basic idea: imagine I have two qubits, where each qubit can be in a $|0\rangle$ state or a $|1\rangle$ state. If I measure both qubits, what are the possible results of my measurement? Hopefully you answered: $|00\rangle$, $|10\rangle$, $|01\rangle$, or $|11\rangle$, where the notation $|00\rangle$ is read, ``qubit \#1 is in state $|0\rangle$, and qubit \#2 is also in state $|0\rangle$." If each qubit has two possible states, then measuring both qubits gives four total possible states. An imperfect analogy: if I have two coins, and I flip each coin, the possible states after flipping are HH, HT, TH, and TT. We can mathematically represent the two-qubit state as:     $$ |\psi\rangle = \alpha |00\rangle + \beta |01\rangle + \gamma |10\rangle + \delta |11\rangle $$
The complex numbers $\alpha$, $\beta$, $\gamma$, $\delta$ have exactly the same interpretation as the numbers $\alpha$ and $\beta$ above: they represent the probability of obtaining each of the above states after measure both qubits. So, for example, $|\gamma|^2 = \gamma^* \gamma$ is the probability of measuring the first qubit to be in the $|1\rangle$ state, and the second qubit to be in the $|0\rangle$ state. I want you to notice something interesting: when we moved from one to two qubits, we doubled the number of complex numbers we need to completely describe the state of our qubits, again denoted by $|\psi\rangle$. If we have three qubits, there will be eight possible outcomes of measuring all three qubits ($|000\rangle$, $|001\rangle$, ..., $|111\rangle$), so we'll need eight complex numbers ($\alpha_1$, $\alpha_2$, ..., $\alpha_8$, where $|\alpha_1|^2$ is the probability of observing $|000\rangle$) to describe the system. Hopefully you're starting to notice a pattern: if we have $n$ qubits in our quantum computer, we need $2^n$ complex numbers to completely describe the state of our quantum computer at any given time. This is manageable if we only have a few qubits, but consider building a quantum computer with 100 qubits (which is about the limit right now for state-of-the-art quantum computers): we'd need $2^{100} \approx 10^{30}$ complex numbers to completely describe this system! In other words, if I wanted a classical computer to accurately simulate a quantum computer with 100 qubits, it would need to store $10^{30}$ complex numbers, which would require about $10^{12}$ times more storage than the entire internet currently uses. Quantum computing is magical because it allows us to access a \textit{state space} that is absolutely massive with only a small number of qubits.

Returning to our hypothetical 100-amino acid polypeptide with about $10^{178}$ possible conformations, let's assume that each conformation can be represented by $2\cdot100 = 200$ numbers (two dihedral angles for each amino acid, 100 amino acids in the chain). This gives $\approx 10^{180}$ numbers to completely describe every possible conformation of the protein. If we could build a quantum computer with 600 qubits, we'd have access to a computational space with $2^{600} \approx 10^{180}$ complex numbers, easily enough to store every possible conformation of the protein! Before getting too excited about this, there are two things worth noting: first, we can't currently build quantum computers with more than about a hundred qubits, and the quantum computers we can build tend to have large error rates. Second, protein folding simulations on quantum computers don't search through every possible conformation, and the quantum algorithms that would be used are extremely complex in both theory and actual implementation. The key point from this conversation is that quantum computers have the computational power to explore \textit{astronomically} more conformations than classical computers, drastically increasing the complexity of systems that we can simulate and predict. Even if this power isn't currently available, focused research and development is exponentially increasing the number of qubits in our quantum computers, so it will hopefully be available soon. 

\textit{(Another technical aside about the above discussion: technically, there are quantum computers with a few thousand qubits produced by a company called D-Wave Systems. However, these quantum computers tend to have very large error rates, and are only designed to perform a very narrow set of optimization calculations. There's also been some controversy over whether their computers actually use quantum effects appropriately \cite{DWAVE}. In the above, I'm referring to research and development quantum computers developed by Google, IBM, NASA, etc., which are not used for commercial purposes and have $\sim 100$ qubits.)}

With this introduction, we can start discussing some papers that are using quantum computing methods to predict protein structures! 

\subsection{Lattice protein folding}

As we mentioned above, quantum computers are currently restricted to $\approx 100$ qubits, and constructing quantum computers that can operate without errors is very difficult. This means that quantum computers cannot currently simulate the folding of a large protein like the AlphaFold neural network. Even with the minimal qubit resources we have available, however, there are problems related to protein folding where quantum computers can offer a significant speedup over classical computers. One of these problems is the \textit{lattice protein folding problem}, which involves predicting the lowest-energy conformation of a protein that is \textit{restricted to a cubic lattice}. A representative paper exploring this area is \cite{ANNEAL}, which focuses on methods to solve the lattice protein folding problem using a state-of-the-art quantum computer. 

We'll define exactly what a lattice protein is, and why it's useful, in just a moment; first, I want to give a roadmap of how \cite{ANNEAL} manages to find the lowest-energy conformation of a lattice protein. This process might not be clear right now, but after reading this section you should have a good conceptual understanding of each part! Here's the procedure: 
    \begin{enumerate}
        \item Generate \textit{constraints} on your lattice protein model, encoding physical laws that your protein should follow (for example, we can't have two residues overlapping in space, so this would be a useful constraint). 
        \item \textit{Encode} these constraints in an energy function similar to $V(\mathcal{C})$ that we discussed above. In this context, the energy function is called a \textit{Hamiltonian}. 
        \item Use \textit{quantum annealing} to minimize this energy function and find a solution to your original problem. 
    \end{enumerate}
I'll walk through each of these steps below, just after I explain what a lattice protein actually is. 

\textbf{Lattice protein models.} In a lattice protein model, we construct a cubic lattice where residues occupy lattice sites and peptide bonds between residues are represented by edges. Figure 1 in \cite{ANNEAL} gives a good visual explanation of this concept; each of the dark grey circles in the diagram represents a residue, while the thick black lines represent the peptide bonds. Notice that at a given residue site (vertex of the lattice), the next peptide bond has at most five possible directions: up/down, left/right, and forward (going backward would result in two superimposed residues, which can't occur). A lattice protein is constructed by placing a residue on a vertex, exploring the possible directions for the next peptide bond, and choosing the direction that (1) results in no residue overlaps, and (2) minimizes the overall energy of the protein. Just from looking at this figure, you should immediately be raising all kinds of objections: ``That's not at all what real proteins look like! You're completely ignoring the side chains! There's not enough detail to accurately reconstruct the entire protein secondary and tertiary structure!" Indeed, these are all excellent objections! It's important to realize that lattice protein models are \textit{not} designed to construct an entire protein, like AlphaFold. Instead, they are designed to complement classical algorithms (that use traditional methods, such as the neural-network models we analyzed earlier) with important information about the \textit{coarse-grained structure} of the overall protein. It turns out that the minimum-energy conformation of a lattice protein can be extremely useful for classical algorithms, but it's extremely difficult\footnote{For the computer-science oriented, it's \textsf{NP}-complete.} for classical computers to actually compute this conformation. However, quantum computers have been shown to solve this particular problem significantly faster than classical computers! 

\textbf{Constraints and the Hamiltonian.} Now that we have the overall model in mind, let's figure out how to actually solve for the lowest-energy conformation. We're once again going to minimize a type of energy function that describes our protein; in this case the energy function is called a Hamiltonian and is denoted by $H$. The notation and name come from quantum mechanics, and the exact origins aren't important for our purposes here. It's instead important to recognize that we can draw very close parallels between $H$ and $V(\mathcal{C})$: both are functions of the protein conformation, and the minimum of both functions corresponds to the lowest-energy state of the protein. We're making a distinction between them because we'll use $H$ in a very different way, and I want to follow the notation in literature. In the following sections, I'll use the terms ``energy function" and ``Hamiltonian" interchangeably, just to emphasize the similarities between $H$ and what we've discussed earlier.  

How do we construct our energy function (Hamiltonian) $H$? Unlike in previous examples, where the energy function was constructed with a neural network or from a probability distribution, the authors of \cite{ANNEAL} construct their energy function ``by hand," using constraints on the lattice protein model that correspond to physical properties of the real-world protein. As an example, we noted above that a lattice peptide bond from residue $i$ has (at most) five possible directions to choose from, because the sixth direction would overlap residues $i + 1$ and $i - 1$. To ensure that this move \textit{increases} the energy function, rather than decreasing it, the authors set up an energy function $H_{\text{back}}$ with the following general structure: 
    \begin{equation*}
        H_{\text{back}} = 
        \begin{cases} 
            \infty, & \text{if} \ r_{i + 1} = r_{i - 1} \\
            0, & \text{if} \ r_{i + 1} \neq r_{i - 1}
        \end{cases}
    \end{equation*}
Here, $r_{i + 1}$ and $r_{i - 1}$ indicate the lattice positions of residues $i + 1$ and $i - 1$. Note that if the two residues overlap, the energy function has an infinite value; since we want to minimize $H_{\text{back}}$, we want non-allowed states (like overlapping residues) to give really big values for our energy function. When we apply our minimization algorithm to find the lowest-energy state, the algorithm will quickly move away from overlapping residues because the corresponding energy is so high. As a note, this is \textit{not} the actual form of $H_{\text{back}}$ that's used in \cite{ANNEAL}, as this form is extremely complicated and would take much more space to explain. I just want us to have a general conceptual idea of how these energy functions are constructed! 

If we have additional constraints to impose, we simply construct additional Hamiltonian terms and sum all of these terms together to obtain an overall Hamiltonian for the lattice protein system. For example, we might require that hydrophobic residues cluster together, since the hydrophobic effect is an extremely important determinant of secondary and tertiary structure. The authors encapsulated this constraint in a term denoted by $H_{\text{pair}}$, which has the general form: 
    \begin{equation*}
        H_{\text{pair}} = 
        \begin{cases} 
            +1, & \text{if} \  d(r_h, r_p) = 1 \\
            -1, & \text{if} \ d(r_h, r_h) = 1 
        \end{cases}
    \end{equation*}
In this case, $r_h$ and $r_p$ represent hydrophobic and polar residues, respectively, and $d(\cdot)$ indicates the distance between the two residues (that is, the number of edges between the two residues). Adjacent hydrophobic residues ($d(r_h, r_h) = 1$) are given a more negative score, while adjacent hydrophobic and polar residues are given a more positive score. Again, this wasn't the exact form used in \cite{ANNEAL}, but it covers the basic conceptual idea! The authors include a few more constraints that you can read about in the paper, and then sum all of these constraints into an overall Hamiltonian $H$:
    $$ H = H_{\text{back}} + H_{\text{pair}} + \cdots $$
Note that minimizing each individual term is equivalent to minimizing the overall Hamiltonian $H$.

\textbf{Quantum annealing.} Once we have a Hamiltonian $H$ that correctly encodes the constraints we've placed on the lattice protein, how do we find the lowest-energy conformation that still satisfies all of the required constraints? The authors of \cite{ANNEAL} used a technique called \textit{quantum annealing} on a quantum computer to solve this problem (see, for example, \cite{ADIABAT}). This technique is based on a theorem in quantum mechanics called the \textit{adiabatic theorem}, and despite the scary-sounding name it's actually pretty simple to describe. Here's the idea: finding the conformation that minimizes the Hamiltonian $H$, called the \textit{ground state} of the Hamiltonian, is a really hard problem to solve. But it's pretty straightforward to design a Hamiltonian where the ground state \textit{is} easy to find, even if this Hamiltonian is useless for the problem we're ultimately trying to solve. We'll call this Hamiltonian $H_{\text{easy}}$, and what's interesting is that it doesn't actually have to be related to the energy function $H$ that we ultimately want to minimize. Aside from a few minor technical constraints that aren't important for our purposes, $H_{\text{easy}}$ just has to be ``easy" to minimize, so we can immediately identify its ground state. For example, we might write 
    $$ H_{\text{easy}} = \sum_{i = 1}^{N} x_i $$
where $x_i \in \{ -1, +1 \}$ and $N \in \mathbb{R}$. Obviously, if we want to make $H_{\text{easy}}$ as small as possible, we just set $x_i = -1$ for all $i \in [1, N]$, so this is the ground state. While this might seem like a simple, contrived example, it's actually pretty close to real ``easy" Hamiltonians that are used in literature! 

When we have an appropriate easy Hamiltonian, what do we do next? This is where we use the \textit{adiabatic theorem} that I mentioned earlier. Suppose I have a quantum computer and an easy Hamiltonian, and I know that the ground state of this easy Hamiltonian corresponds to the ``all $|0\rangle$" state of the quantum computer. In other words, I know that setting each qubit equal to $|0\rangle$ minimizes the energy function $H_{\text{easy}}$ that I've designed, so the quantum computer is in the ground state of this Hamiltonian. Suppose I then construct the following energy function: 
    $$ H_a(t) = (1 - t/\tau) H_{\text{easy}} + (t/\tau) H $$
where $H_a(t)$ stands for ``adiabatic Hamiltonian," $t$ is time, $\tau$ is a constant, and $H$ is the Hamiltonian for the protein we want to solve. Note that for $t = 0$, we just have $H_a(0) = H_{\text{easy}}$, but for $t = \tau$ we have $H_a(\tau) = H$. The adiabatic theorem states that if I perform operations on my quantum computer (which is in the ground state of $H_{\text{easy}}$) slowly enough over some time $\tau$, I can ``transform" the state of my quantum computer from the ground state of $H_{\text{easy}}$ to the ground state of $H$. Here's another way to think about it: as $t \rightarrow \tau$, the ``easy" energy function goes to zero, and I'm left with the ``hard" energy function $H$. If I move to the hard Hamiltonian slowly enough, the state of my quantum computer is ``pulled along" until it's also in the ground state of $H$, instead of in the ground state of $H_{\text{easy}}$. Since we designed the ground state of $H$ to encode the minimum-energy conformation of the protein, our quantum computer now effectively stores a representation of this conformation. If we have some way to decipher this quantum state, we can obtain the conformation and solve our problem. 

This might seem like magic, and to be honest it pretty much is, since basically everything in quantum mechanics feels like magic. Of course, the details of implementing a system like this are extremely complicated: we have to make sure that our Hamiltonian is designed in a way that makes it easy to ``extract" information about the protein conformation from the final state of our quantum computer, and our quantum computer has to perform the calculation without significant errors. There are also significant restrictions on how quickly we can perform operations to shift from $H_{\text{easy}}$ to $H$, as moving too fast causes the procedure to fail. Despite these difficulties, quantum computers today can perform this type of calculation, and it's exactly the technique used by the authors of \cite{ANNEAL} to solve for the minimum-energy conformation of the lattice protein. It's also important to emphasize that the adiabatic theorem, and the overall technique of quantum annealing using the adiabatic theorem, has no classical analog, so this type of calculation can only be performed on a quantum computer. 

\textit{(A note on nomenclature: quantum annealing is the process of using the adiabatic theorem to solve for the ground state of some ``hard" Hamiltonian.)}

\begin{figure}
    \centering
    \includegraphics[scale=0.5]{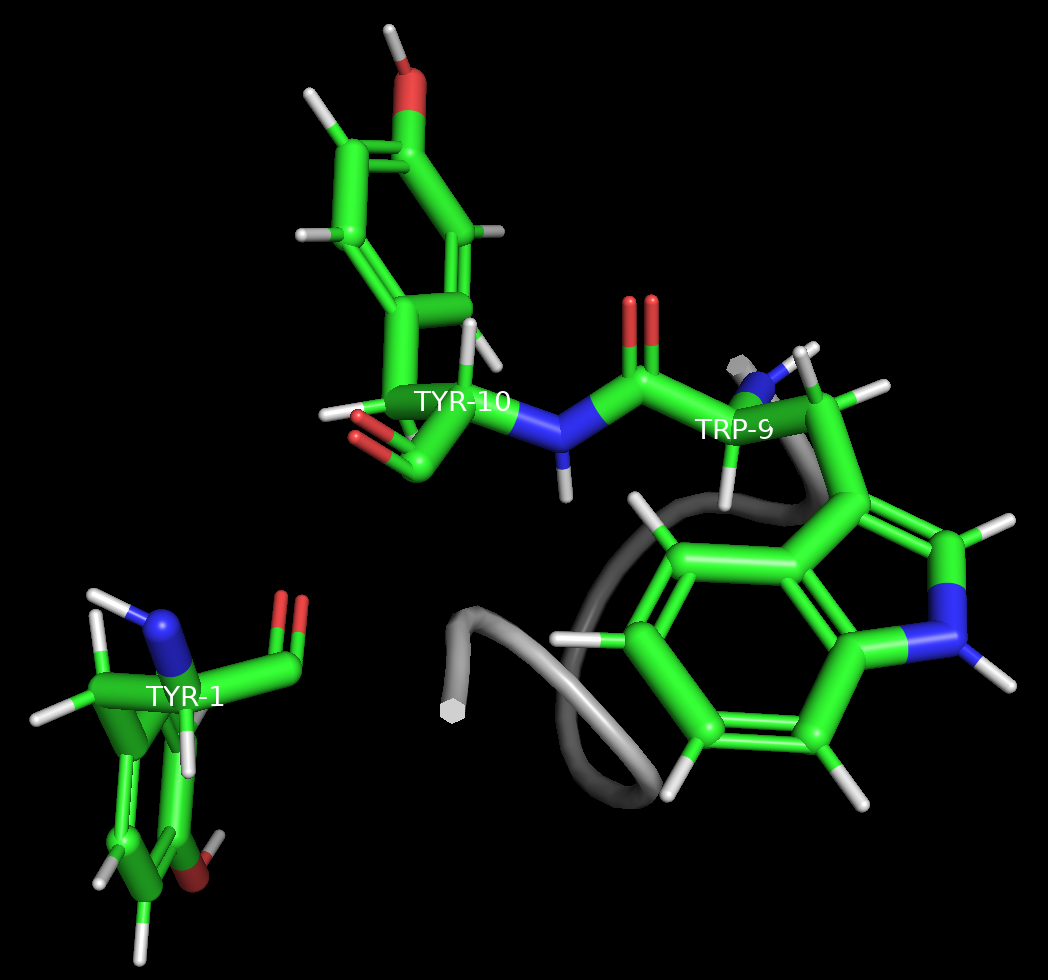}
    \caption{The $\text{Y} - \text{Y}$ (Tyr1 $-$ Tyr10) and $\text{Y} - \text{W}$ (Tyr10 $-$ Trp9) contacts successfully predicted for the 10-residue Chignolin protein. Note that the side chains are highly conjugated, presumably leading to energy-lowering $\pi-\pi$ interactions. The remaining 7 residues are represented by the gray line in the background, for clarity.}
    \label{yyw}
\end{figure}

The authors of \cite{ANNEAL} used the quantum annealing process described above to solve the lattice structures of two proteins, Chignolin (10 residues; PDB 2RVD) and Trp-Cage (20 residues, although they solved a fragment of 8 residues; PDB 2JOF). They claim to be the first group to successfully fold a lattice protein on a three-dimensional lattice using a quantum computer, as previous work solved the same problem only on two-dimensional lattices. An interesting note on their results is that they were successfully able to predict inter-residue interactions within the protein, based on the topology of the final conformation. For example, when folding the lattice model of the Chignolin protein, their model predicted $\text{Y}-\text{Y}$ and $\text{Y}-\text{W}$ interactions based on the distance between these residues in the lattice (see figure \ref{yyw}). Both of these contacts are formed in the actual Chignolin protein, and the authors note that this type of contact information for an unknown protein could be highly beneficial for classical algorithms like AlphaFold.

Clearly the technology is still in its infancy, as current quantum computers can only handle proteins or protein fragments with tens of residues. Given the exponential increasing in quantum processing power, however, I anticipate that the number of residues in solved lattice proteins will increase substantially over the next several years! 

\section{Conclusions}

Let's take a step back and review what we've covered: starting with Anfinsen's classical experiments on protein structure in the 1960's, we were able to explore the thermodynamic hypothesis and Levinthal's paradox to set the stage for modern computational protein folding. We then looked at some current state-of-the-art methods, focusing first on machine learning algorithms that can elucidate new structures after being trained on thousands of previously solved proteins. We saw that these algorithms can be neural-network focused, so most of the physics of the problem is abstracted to the network's training, or can be physics-based, so equations and concepts from statistical physics (for example, the Boltzmann distribution) can guide the algorithm's problem-solving. We then moved to quantum computing methods, providing a brief primer on basic concepts in quantum information and analyzing why quantum computing could be useful for folding proteins. We finally analyzed a quantum annealing model that was able to fold a lattice protein and faithfully reproduce inter-residue contacts that are present in the protein's native structure. 

While this certainly doesn't come close to encapsulating all of the research and methods that have been used for protein folding, I'm hoping you now have a better idea of how computational protein folding has developed and become excited about future possibilities! As we mentioned in the super pretentious introduction, protein folding is incredibly important for nearly every aspect of medicine, especially drug design and the detection and treatment of cancers. Developing fast and accurate tools to fold proteins successfully could save millions of lives in future decades, and the next great idea could come from you. So go take a physics class, or a structural biology class, or go to medical school, or just read more about protein folding in medicine and how it might relate to your field of study. I can't wait to see what you come up with! 

\section{Acknowledgements}

A big thanks to the Chemistry 143 teaching team at Stanford for their support; this paper was originally written as a final project for this class during the spring of 2020. 

\textbf{Note:} Every effort has been made to ensure the accuracy and clarity of the information presented in this article. However, if you find a mistake or explanation that isn't clear, please feel free to contact me at \url{smullane@stanford.edu} so the article can be updated for future readers. Thank you! 

\newpage

\bibliographystyle{plain}
\bibliography{ref_update}

\end{document}